\documentclass[aps,twocolumn,prd,showpacs,nofootinbib]{revtex4}
\usepackage{amsmath}
\usepackage{graphicx}
\usepackage{dcolumn}
\usepackage{bm}
\usepackage{amssymb}
\usepackage{latexsym}

\def\be{\begin{equation}}
\def\ee{\end{equation}}
\def\ba{\begin{eqnarray}}
\def\ea{\end{eqnarray}}

\begin{document}

\title{ Seeding of Primordial Perturbations
During a Decelerated Expansion }

\author{Yun-Song Piao}
\affiliation{College of Physical Sciences, Graduate School of
Chinese Academy of Sciences, YuQuan Road 19{\rm A}, Beijing
100049, China} %\affiliation{$^b$CCAST (World Lab.), P.O. Box 8730,
%Beijing 100080}
%\affiliation{Institute of Theoretical Physics,
%Chinese Academy of Sciences, P.O. Box 2735, Beijing 100080, China}

\begin{abstract}

A scalar field with a modified dispersion relation may seed, under
certain conditions, the primordial perturbations during a
decelerated expansion. In this note we examine whether and how
these perturbations can be responsible for the structure formation
of observable universe. %We find that there seems slightly
%difficult in matching the requirements of observable cosmology, i.e. obtain
%simultaneously the scale invariant spectrum of primordial
%perturbations and enough ``efolding number", in a simple case.
%however, when
% with the
%massless scalar field. However, when
%the mass of scalar field is introduced, the problem will not
%exist.
%it is still possible to
We discuss relevant difficulties and possible solutions.

\end{abstract}

 \maketitle

\section{introduction}

Recently, lots of observations have payed attention to the nature
of primordial perturbations that gave rise to the inhomogeneities
observed in the universe. The results of these measurements are
consistent with an adiabatic and nearly scale invariant spectrum
of primordial perturbations, as predicted by the simplest models
of inflation \cite{WMAP}. However, due to the central role of
primordial perturbations on the formation of cosmological
structure, it is still very interesting and might also be
desirable to explore various and possible origins of primordial
perturbations.

The inflation stage is supposed to have taken place at the earlier
moments of the universe \cite{Guth, LAS}, which superluminally
stretched a tiny patch to become our observable universe today,
and in the meantime makes the quantum fluctuations in the horizon
leave the horizon to become the primordial perturbations
responsible for the formation of cosmological structure
%when their reentering into the
%horizon after the end of inflation
\cite{BST, MC}. This is one of remarkable successes of inflation,
%However, it can be noted that this generation of superhorizon
%primordial perturbations can actually also be implemented in the
%contracting background \cite{S, VGV, KOS, FB}, and
see also the superinflation e.g. Refs. \cite{PZ, PZ1, GJ, BFM}, in
which the null energy condition is broken. In Ref. \cite{PZ}, it
was firstly noticed that there is an interesting case in the
generating phases of primordial perturbations, in which the scale
factor and thus the wavelengths of perturbations grows very slowly
but the Hubble length rapidly shrinks. The inflation can be
generally regarded as an accelerated or superaccelerated stage,
and so may defined as an epoch when the comoving Hubble length
decreases. This length starts out very large, and then the
inflation forces it to shrink enough so that the perturbations can
be generated causally. In a decelerated expanding background the
comoving Hubble length is increased, thus in this case it seems
hardly possible to causally explain the origin of primordial
perturbations. The variable speed of light \cite{M, AM} has been
considered, however, also see Ref. \cite{EU} for a reexamination.
Note that it has been illustrated in Ref. \cite{Liddle} that the
existence of adiabatic perturbations on scales much larger than
the Hubble radius implies that either inflation occurred in the
past, the perturbations were there as initial conditions, or
causality is broken. %However, if we expect that the
%primordial perturbations result from the quantum fluctuations of
%the scalar field but not thermal fluctuations \cite{MP}, the above
%no-go theorem is validated only in usual case in which the scalar
%fields have the normal dispersion relation.
Thus if we want to obtain the primordial perturbations in a
decelerated expanding phase, we have to require that the scalar
field responsible for the perturbations should have some special
or modified dispersion relation. %These modifications under certain
%conditions may provide an interesting seeding mechanism
%of primordial perturbations. % which
%can be expected from possible high energy correction of physical
%laws in which the spatial momentum of matter perturbations can be
%changed.
%It relies on a modified
%dispersion relation that contains higher powers of the spatial
%momentum of matter perturbations, which
Though there have been many detailed descriptions how the required
dispersion relations are obtained from the effective field theory
\cite{Picon}, it will be still significant to examine the
feasibility of this seeding mechanism matched to the observable
cosmology.
%We find that in
%order to generate a scale invariant spectrum, it will be slightly
%difficult to obtain enough efolding number required by observable
%cosmology for a massless scalar field. We also discuss possible
%solution to this problem.
%To clearly understand the generation of primordial perturbations
%of scalar field with the modified dispersion relation, let us
%firstly see that with the normal dispersion relation in the
%accelerated expanding background.

The outline of this paper is as follows. In section II, we will
show how the primordial perturbation may be generated during a
decelerated expanding phase when the dispersion relation is
modified. In this case, the perturbation spectrum is calculated in
section III. We discuss relevant difficulties in matching the
spectrum to observable cosmology and possible solutions. Finally,
we summarize and discuss our results, and as well as give some
comments on future issues.

\section{generation of spectrum}

In this section we will begin with a general discussion on the
generation of causal primordial perturbations. The generation of
primordial perturbations requires that the perturbation modes can
leave the horizon during their generation and then reenter the
horizon at late time. Thus it may be convenient to define \be
{\cal N} \equiv \ln\left({k_e\over k}\right)\equiv \left({a_e h_e
\over a h}\right), \label{caln}\ee which measures the efolding
number of mode with some scale $\sim k^{-1}$ which leaves the
horizon before the end of the generating phase of perturbations,
see Ref. \cite{KST}, where $k$ is the comoving wave number, and
the subscript `$e$' denotes the end time of the generating phase
of perturbations, thus $k_e$ is the last mode to be generated, and
$h\equiv {\dot a}/a$ is the Hubble parameter, where the dot
denotes the derivative with respect to the cosmic time. When
taking $ah=a_0h_0$, where the subscript `0' denotes the present
time, we will obtain the efolding number required by observable
cosmology. In this case, Eq.(\ref{caln}) is actually the ratio of
the physical wavelength corresponding to the present observable
scale to that at the end of the generating phase of perturbations.

The evolution of scale factor in the expanding background can be
simply taken as $a(t) \sim t^n$, ($t\rightarrow \infty$) and $
a(t) \sim (-t)^{n}$, ($t\rightarrow 0_-$). We will assume that $n$
is a constant for simplicity.
%, and work
%with the parameter $\epsilon \equiv-{\dot h}/h^2$, which describes
%the change of $h$ in unit of Hubble time and depicts the abrupt
%degree of background change.
%For above background we have $\epsilon= 1/n < 1$.
In the conformal time $\eta$, we obtain $ a(\eta) \sim
(-\eta)^{n/(1-n)}$. Thus we have \be a\sim \left({n\over (n-1) a
h}\right)^{n\over 1-n}.\label{ah}\ee
%\be h \equiv {a^\prime \over a^2}=
%{1\over (\epsilon -1) a \eta}, \label{htau}\ee
To produce the efolding number, i.e. ${\cal N}>0$, $ah$ must
increase with time, i.e. ${\ddot a}>0$. This suggests that $n>1$
for $a(t)\sim t^n$, which corresponds to the accelerated expanding
phases, in which ${\dot h}<0$, and $n<0$ for $a(t)\sim (-t)^n$,
which corresponds to the superaccelerated expanding phases, in
which ${\dot h}>0$, e.g. see Ref. \cite{PZ2} \footnote{The limit
case of $n\simeq 0_-$ can be very interesting \cite{PZ} and was
studied in detail in the island universe model \cite{Piao,
Piaoii}.}.
%which is just the value
%of $n$ for above accelerated expanding background.
%From Eq.(\ref{ah}), we can obtain \be {a_e\over a}=\left({a h\over
%a_e h_e}\right)^{n\over 1-n} = \left({k\over k_e}\right)^{n\over
%1-n} , \label{asim}\ee which, combined with Eq.(\ref{caln}),
%induce $a_e/a= e^{n{\cal N}/( n-1)}$. Thus we can see that when
%$n\rightarrow \infty$, the scale factor exponentially expands,
%which is the case of usual inflation models, while when $n<0$, the
%change of scale factor can be very small. In the limit $n\simeq
%0_-$, the scale factor will be nearly unchanged, which corresponds
%to the case in Ref. \cite{Piao, Piaoii}. From Eqs.(\ref{caln}) and
%(\ref{asim}), we can obtain \be {h_e\over h} =\left({a_e h_e\over
%a h}\right)^{1\over 1-n}= \left({k_e\over k}\right)^{1\over 1-n},
%\label{hhe}\ee which will be used in the following.
%We can see that when $n\rightarrow \infty$, $h$ is nearly
%unchanged, which is the case of usual inflation models.
%while when $n\simeq 0_-$, we have $h_e/h\simeq
%k_e/k$, i.e. the efolding number. %The above descriptions are also
%seen as follows.
Taking the logarithm in both sides of (\ref{ah}), we obtain \be
\ln{\left({1\over ah}\right)}=\left({1-n\over n}\right)\ln{a} .
\label{ahi} \ee This equation can be also applied to the expansion
with
arbitrary constant $n$. %To illustrate the characters of those
%(super)accelerated expanding phases able to generate the
%primordial perturbations,
We plot Fig.1, in which and also in the whole note we have assumed
that after the generating phases of primordial perturbations ends,
the ``reheating" will rapidly occur and then bring the universe
back to the usual FRW evolution \footnote{The superaccelerated
phase will generally evolve to a Big Rip at late time, unless
there is not an exit mechanism or ``reheating".}. Here the
reheating means a transition that the fields or fluids dominating
the background decay into radiation.

%\begin{figure}[t]
%\begin{center}
%\includegraphics[width=6cm]{decelei.eps}
%\caption{The figure of evolution of $\ln{(1/ah)}$ with respect to
%the scale factor $\ln{a}$ during the generation of perturbations
%for different $n$. The left side of $a_e$ is the generating phase
%of perturbations, in which the yellow line denotes the usual
%exponential expansion with $|n|\rightarrow \infty$, which divides
%the region into the superaccelerating phase (up right) in which
%$n<0$ and the accelerating phase (lower left) in which $n>1$. The
%blue, tint blue and red lines denote the various
%(super)accelerating phases with $n\simeq 0_-, -1, 2$,
%respectively. The perturbation modes with the wave number $k$ can
%leave the Hubble horizon during their generations and then reenter
%the horizon during the radiation and matter -dominated at late
%time. In principle it can hardly possible to obtain the
%superhorizon primordial perturbations in a decelerated expanding
%phases in which $0<n<1$, since there are nothing leaving the
%horizon during their evolutions. }
%\end{center}
%\end{figure}

\begin{figure}[t]
\begin{center}
\includegraphics[width=7cm]{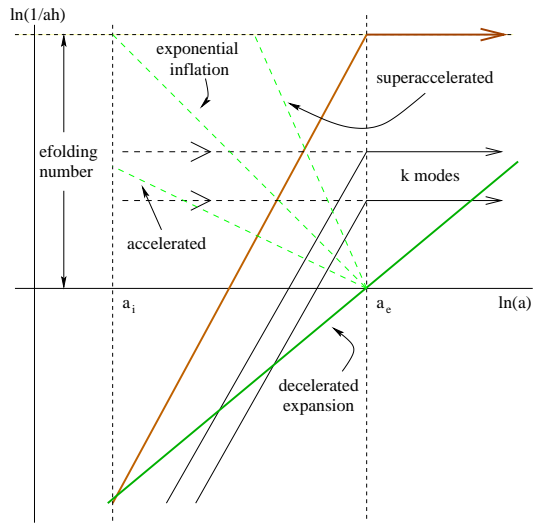}
\caption{The sketch of evolution of $\ln{(1/ah)}$ with respect to
the scale factor $\ln{a}$ during the generation of perturbations.
The left side of $a_e$ is the generating phase of perturbations,
in which in the region above the $\ln{a}$ axis the green dashed
line corresponding the usual exponential inflation with
$|n|\rightarrow \infty$ divides this region into the
superaccelerated phase (up right) in which $n<0$ and the
accelerated phase (lower left) in which $n>1$. The perturbation
modes (black dashed lines) with the wave number $k$ can leave the
Hubble horizon during their generations and then reenter the
horizon during the radiation/matter domination at late time. In
principle it is hardly possible to obtain the causal primordial
perturbations in a decelerated expanding phases in which $0<n<1$,
see the region below the $\ln{a}$ axis, since there are nothing
leaving the horizon during their evolutions. However, when we
introduce the scalar field with the modified dispersion relation
$\omega=k/a^p$, the similar case to inflation can be imitated in a
decelerated expanding phases. In this case the comoving wave
number of perturbations will not unchange any more. This makes the
evolution of their physical wavelengths able to be faster than
that of $1/h$ when the condition $n(p+1)-1>0$ is satisfied, see
the black line and green line. }
\end{center}
\end{figure}

%From the above descriptions and Fig.1,
We can see in Fig.1 that in principle one can not obtain the
primordial perturbations in the decelerated
expanding phases in which $0<n<1$, however, %it has been shown in Ref.
%that
introducing the modified dispersion relation can change this
point. The usual dispersion relation may be generally expected to
receive some corrections with the increase of energy, which in
some sense can be regarded as a phenomenological description of
high energy new physics, see Ref. \cite{U, CJ}. These
modifications has been applied to the early universe, especially
the inflation cosmology, in which the modified dispersion relation
can significantly affect the spectrum of primordial perturbations
generated during inflation \cite{MB, BM1}. The modified dispersion
relation can also naturally arise from generally covariant scalar
field \cite{JM}. Following Ref. \cite{Picon}, in the conformal
time, one can introduce a projector $h_{\mu\nu}$, which projects
onto the space orthogonal to a timelike vector $u_\mu=(1/a,0,0,0)$
and satisfies $h_{\mu\nu}h^{\nu}_{\,\rho}=h_{\mu\rho}$ and
$h_{\mu\nu}u^{\nu}=0$. Thus one may write $h_{\mu\nu}$ as
$h_{\mu\nu}=a^2\cdot (0,1,1,1)$. By the help of the projection
tensor, we can define a spatial derivative as ${\cal
D}_{\mu}=h_{\mu}^{\nu}\nabla_{\nu}$, which is orthogonal to
$u_{\mu}$ and so has only spatial components, and a time
derivative $(\delta_{\mu}^{\,\nu}-h_{\mu}^{\,\nu}) \nabla_{\nu}$.
In principle, by combining these two generally covariant
derivatives, one can obtain any combination of time and spatial
derivatives, so any dispersion relation when acting on a scalar
field. For example, if the spatial component of scalar field
Lagrangian is as follows $\sim\varphi ({\cal D}_{\mu}{\cal
D}^{\mu})^q \varphi/ a^p $, where $p$ and $q$ are constant, we
will have the dispersion relation \be \omega= {k^q\over
a^p}.\label{omega}\ee Here for our purpose we will not pay more
attentions to the relevant discussions on the modified dispersion
relation. We in the following will focus on the primordial
perturbations of the scalar field with the dispersion relation
(\ref{omega}) in a decelerated expanding background.

We firstly begin with a simple modification as follows $\omega =
k/a^{p}$, where $p$ is the constant, and when $p=0$, it recovers
to the normal one. Note also that this case in some sense is
similar to that of the fluid with the decaying sound speed
\cite{AM1, Picon2}. To make a matching to the observable
cosmology, it is convenient to define an equivalent ``efolding
number". Note that with above modification to the dispersion
relation, the effective comoving wave number $\omega$ will not
unchange any more during its evolution and has an extra suppress
leaded by $a^{p}$, and thus an increasing $\sim a^{p}$ of the
effective comoving wavelength, which directly effects the physical
wavelength of correspondent mode. What the efolding number
required by observable cosmology actually reflects is the ratio of
the physical wavelength corresponding to the present observable
scale to that at the end of the generating phase of perturbations,
thus the change of physical wavelength induced by the shift of the
comoving wave number must be included in the definition of the
efolding number. Thus similar to Eq.(\ref{caln}), the equivalent
efolding number can be written as \be
%{\cal N} \equiv \ln\left(\left({a_e\over a}\right)^p\cdot
%{k_e\over k}\right)
e^{\cal N}\equiv \left({a_e\over a}\right)^p \cdot \left({k_e\over
k}\right) .
%\equiv \left({a_e\over a}\right)^p \cdot \left({a_eh_e\over
%ah}\right)
\label{caln2}\ee From Eq.(\ref{ah}), we have $
\ln{(k_e/k)}=(n-1)\ln{(h/ h_e)} $. Thus substituting it and
Eq.(\ref{ah}) into Eq.(\ref{caln2}), we can obtain
%\be {\cal N} = \left(np+p-1\right)\ln\left({h\over
%h_e}\right) .\label{caln3}\ee
${\cal N} = (np+n-1)\ln{(h/ h_e)} $. For an expanding universe
with $0<n<1$, we can generally have $h>h_e$. Thus to make ${\cal
N}>0$, which is required by the generation of primordial
perturbations, $n(p+1)-1>0$ must be satisfied. This can be reduced
to $n>1$ for $p=0$, which
corresponds to the usual inflationary cases. %To illustrate the
%essential of generation of primordial perturbations in the scalar
%field with modified dispersion relation, we plot Fig.2.
In the expanding process, for the field with the normal dispersion
relation, the physical wavelength of its modes $\sim a$, and only
when the evolution of $a$ is faster than that of $1/h$, can the
primordial perturbations be generated, which can only be
implemented in the cases of $n>1$ and $n<0$ (here $h<h_e$).
However, for the field with the modified dispersion relation $
k/a^p$, the physical wavelength of correspondent modes is $\sim
a\cdot a^p$, thus even if for $0<n<1$ it is also possible that the
evolution of physical wavelength is faster than that of $1/h$,
which leads that the correspondent modes can leave the horizon and
thus the generation of primordial perturbation, see Fig.1 for an
illustration. In fact the condition $n(p+1)-1>0$ means that in
Fig.1 the lines of $k$ modes must be intersected with that of
$\ln{(1/ah)}$, i.e. the green line, in the past. If
$n(p+1)-1\simeq 0$, the intersection will be expected to be in
infinite far position of lower left side of Fig.1, and in this
case the scale of $h$ will be required to be very high. Note that
$h$ can be taken to Planck scale at most and in principle $h_e$
has also a lower limit, thus generally $n(p+1)-1$ should be far
away from 0 in order to obtain enough efolding number.

\section{calculations of spectrum }

In this section we will calculate the primordial perturbations
spectrum of scalar field with modified dispersion relation. We
assume that this scalar field $\varphi$ dose not affect the
evolution of the background. In the momentum space, the motion
equation of $\varphi$ is given by \be u_k^{\prime\prime}
+\left(\omega^2-f(\eta)\right) u_k = 0 ,\label{uk}\ee where $u_k$
is related to the perturbation of $\varphi$ by $u_k \equiv a
\varphi_k$ and the prime denotes the derivative with respect to
$\eta$, and $\omega$ is given by Eq.(\ref{omega}) with $q=1$, and
generally since $\omega=k/a^p\sim k/(-\eta)^{n\over 1-n}$, which
is different from the usual case with $\omega=k$ constant, thus by
using the mathematics handbook about deformed Bessel equation,
$f(\eta)$ is required to be written as \be f(\eta)\equiv
{v^2r^2-1/4 \over \eta^2}, \label{feta}\ee where $v$ is generally
required to be nearly constant so that Eq.(\ref{uk}) is solvable,
and is determined by the evolution of background and the details
of $\varphi$ field, such as its mass, its coupling to the
background, and $r$ is determined by the behavior of
$\omega(\eta)$. Conventionally, the dispersion relation of scalar
field is $\omega=k$, which corresponds to $q=1$ and $p=0$, and
thus $r=1$. When $p\neq 0$, we have \be r\equiv {n(p+1)-1\over
n-1}. \label{r}\ee
%We have, from (\ref{aeta}), \be
%{a^{\prime\prime}\over a}= {\nu^2-{1\over 4}\over \eta^2} \ee
%where $\nu = |1/ 2-1/(\epsilon -1)|$.
%given by \be f_k = \sqrt{-k\eta}(c_1(k){\cal
%H}_{\nu}^{(1)}(-k\eta)+c_2(k){\cal H}_{\nu}^{(2)}(-k\eta)) \ee
%where ${\cal H}_{\nu}^{(1)}$ and ${\cal H}_{\nu}^{(2)}$ are the
%Hankel functions of the first and second kind, respectively, and
%the functions $c_i(k)$ can be determined by specifying the initial
%conditions.
The general solutions of this equation are the Hankel functions
with the order $v$ and the variable $\omega\eta$. In the regime
$\omega\eta \gg 1 $, the modes can be regarded as adiabatic. The
reason is that note that $\omega^{\prime}/\omega \sim 1/\eta$,
thus the adiabatic condition $\omega^{\prime}/\omega^2\ll 1$ is
equivalent to $\omega\eta \gg 1 $. Note also that $\eta\sim
1/(ah)$, we have $\omega\eta\sim \omega/(ah)\gg 1$, and thus
obtain $a\omega^{-1} \ll 1/h$, which corresponds to the case that
the effective physical wavelength is very deep into the horizon.
Thus in this regime we may take \be u_k\simeq {1\over
\sqrt{2\omega(k,\eta)}}\exp{(-i\int^{\eta}\omega(k,\eta)d\eta)}
\ee as an approximate solution of Eq.(\ref{uk}), which in some
sense is similar to the case in which the initial condition can be
taken as usual Minkowski vacuum.
%Thus
%the exact solution becomes \be f_k= {\sqrt{\pi}\over 2 }
%e^{i(\nu+{1\over 2}){\pi\over 2}}\sqrt{-k\eta} {\cal
%H}_{\nu}^{(1)}(-k\eta) \ee
Note that $\omega\eta $ will decrease with the expansion. Thus at
late time, we can expect $\omega\eta \ll 1$, i.e. $a\omega^{-1}\gg
1/h$. The expansion of Hankel functions to the leading term of
$\omega\eta$ gives \be k^{3/2}|\varphi_k | \sim k^{3/2-v}
%e^{i(\nu-{1\over 2}){\pi \over
%2}}2^{(\nu-{3\over 2})} {\Gamma(\nu)\over \Gamma(3/2)}
%{1\over \sqrt{2k}}(-k\eta)^{{1\over 2}-v}
,\label{kuk}\ee where the other factors without $k$ have been
neglected.
% and $v$ is a nearly constant and can be determined by
%the evolution of background and the details of $\varphi$ field,
%such as its mass, its couple to the background.
Thus to obtain the scale invariant spectrum, $v=3/2$ is required.

\subsection{massless case}
For the massless scalar field, we have $f(\eta)\equiv
a^{\prime\prime}/a$. Note that $a\sim \eta^{n/(1-n)}$, and then
use Eqs.(\ref{feta}) and (\ref{r}), we can obtain \be v={1\over
2}\left|{3n-1\over n(p+1)-1}\right|. \label{v}\ee
%where
%the subscript `l' denotes the massless scalar field.
If $p=0$, Eq.(\ref{v}) will be reduced to the normal case, in
which when $|n|\rightarrow \infty$ we can obtain $v=3/2$ and thus
a scale invariant spectrum, which is familiar result in the
inflation. From above discussions we have known that $n(p+1)-1$ is
require to be larger than $0$ for the generation of primordial
perturbations. Thus in Eq.(\ref{v}) taking $v= 3/2$ required by
the scale invariant spectrum, we have \be n(p+1)-1=
\left|n-{1\over 3} \right|.\ee
%for $1/3<n<1$,
%we have $mn = 2/3$ and for $0<n<1/3$, $mn = -2n + 4/3$. Then
Substituting it to ${\cal N}$ leaded by Eq.(\ref{caln2}), we
obtain
%${\cal N}=|n-1/3
%|\ln(h/h_e)$.
\be {\cal N}=\left|n-{1\over 3} \right|\ln\left({h_i\over
h_e}\right), \label{nl}\ee
% and ${\cal N}=(1/3-n)\ln(h/h_e)$ respectively.
where the subscript `$i$' denotes the beginning time of the
generating phase of perturbations. Note that in Eq.(\ref{nl}), the
efolding number is not related with $p$, which is also actually
valid for the case with arbitrary spectrum, though we are
constrained to that with the scale invariant spectrum and then
obtain Eq.(\ref{nl}). Thus since $0<n <1$, we see that to obtain
enough efolding number required by observable cosmology, the ratio
$h_i$ to $h_e$ must be large enough. Note that $h_i$ has an upper
limit, i.e. the Planck scale, thus it seems that $h_e$ must be
taken very low.

The efolding number required is generally determined by the
evolution after reheating. In principle the lower energy density
is when the generating phase of perturbations is over, the smaller
the efolding number required is. For an idealistic case, in which
after the generating phases of perturbations ends the universe
will rapidly be linked to an usual FRW evolution, one can have $
{\cal N} \simeq 68.5+ (1/2)\ln(h_e/m_p)$ \cite{LL}. Instituting it
into (\ref{nl}), we can cancel ${\cal N}$ and obtain a relation
between $h_e$ and $n$. We plot Fig.2 in which for various $n$ in
the region $0.4<n<1$, the $\log (h_i/h_e) $ required are given. We
can see that when taking the initial energy scale as Planck scale
and the end scale as nucleosynthesis scale, in which we have
$h_i/h_e\sim 10^{40}$, in order to obtain enough efolding number,
$n>0.6$ is required, while when the end scale lies in Tev scale,
we have $n>0.8$.
%we can obtain ${\cal N}\simeq 92|n-{1\over 3}|$, and thus $n>0.88$
%for ${\cal N}>50$.
We may also consider slightly red spectrum i.e. $v> 3/2$, as was
favored mildly by WMAP \cite{WMAP}. From Eqs.(\ref{caln2}) and
(\ref{v}), we see that the value of $n$ will be required to be
larger. These results indicate that generally it seems slightly
difficult to satisfy simultaneously the conditions required by the
enough efolding number and the scale invariant spectrum, since the
$h_e$ required must be very low and in the meantime $n$ is
generally constrained in a cabined region. This makes some
familiar phases, e.g. the radiation phase and the matter phase,
hardly be included in possible applications.

%In principle there is an uncertain of the value of the required
%efolding number dependent of the cosmological evolution at late
%time \cite{LL}. If we require $N>60$, we will obtain $n>1$, thus
%the case will be worsen for larger ${\cal N}$.

We may introduce the more general dispersion relation, e.g. change
the power of $k$, as in Eq.(\ref{omega}) with $q\neq 1$. However,
it seems not helpful to relax the above difficult, since this only
equals to bring a factor proportional to $1/q$ in ${\cal N}$
obtained from Eq.(\ref{caln2}) and the denominator of Eq.(\ref{v})
simultaneously. They will be generally set off in the calculations
obtaining the Eq.(\ref{nl}). Thus to solve the above problem, we
must assure that the modifications introduced dose not change
${\cal N}$ obtained from Eq.(\ref{caln2}) and the denominator of
Eq.(\ref{v}) simultaneously.

\begin{figure}[t]
\begin{center}
\includegraphics[width=8cm]{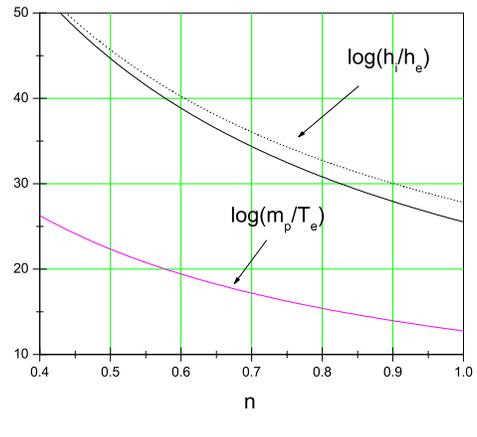}
\caption{The value of the $\log (h_i/h_e) $ with respect to $n$ in
order to obtain enough efolding number. %The vertical axis is $\log
%(h_i/h_e) $ and
The horizonal axis is $n$. The solid line is the case of $h_i\sim
m_p$ and the dashed line is that of $h_i\sim 10^{-4}m_p$. The
region above the corresponding line is that with enough efolding
number. The red line corresponds to the maximal reheating
temperature at $h_e$ epoch where $h_i$ is taken as Planck scale. }
\end{center}
\end{figure}

\subsection{massive case}
When the scalar field is massive, $f(\eta)$ is given by
$f(\eta)={a^{\prime\prime}\over a}- \mu(ah)^2$,  where $\mu\equiv
{m^2_\varphi \over h^2} $ has been defined, see Ref. \cite{PZ3}
for details on the spectrum of massive scalar field with normal
dispersion relation. Note that $a\sim \eta^{n/(1-n)}$, thus we can
obtain $f(\eta)$, where $ah(n-1)/n =\eta$ has been used. Note that
$h\sim 1/t$, i.e. it generally changes with the time in the
decelerated expanding phase, thus $\mu$ is generally not constant
and so the numerator of $f(\eta)$, which will make us very
difficult to obtain the analytic solution of Eq.(\ref{uk}). Thus
we need to fix $\mu$ constant, which can be done by
%solution., which will makes
introducing a non-minimally coupling $\sim R\varphi^2$ between
$\varphi$ and gravity, where $R \sim h^2$ is the Ricci curvature
scalar. In this case, we will have that $m^2_\varphi \sim R\sim
h^2 $, and so $\mu$ can be a constant. Thus Eq.(\ref{uk}) becomes
%the Bessel equation with some deformation and so
solvable exactly.
%$f(\eta)\equiv a^{\prime\prime}/a
%-m_{\varphi}^2 a^2$,
%Note that if $m_{\varphi}$ is far larger than $h(1-\epsilon)$,
%$v_m$ will become imaginary, and thus can not be attached to the
%spectrum index of $\varphi_k$, since in order to obtain the power
%spectrum of $\varphi$ we need to take the module square of $u_k$
%in Eq.(\ref{uk}). Thus the value of $m^2_{\varphi}$ should be
%small positive, or negative for our purpose.
%Thus generally Eq.(\ref{uk}) will be not the exact Bessel equation
%any more, which will make us very difficult to obtain its analytic
%solution. However, there are also some exceptions. When making the
%mass term change with the time as follows $m_{\varphi}^2 a^2\sim
%1/\eta^2 $, which in fact can be motivated by a non-minimally
%coupling of $\varphi$ to gravitation $\sim R\varphi^2$, where $R
%\sim 1/(a\eta)^2$ is the Ricci curvature scalar, we can obtain
%that $m^2_\varphi /h^2 \equiv \mu$ is approximately a constant.
From $f(\eta)$, and then using Eqs.(\ref{feta}) and (\ref{r}), we
can obtain \be v^2_m=v^2-\left({n\over n(p+1)-1 }\right)^2
\cdot\mu
%{m_\varphi^2\over h^2}}
\label{v2}\ee
%$2v=\sqrt{((3n-1)^2-(2n)^2({m_{\varphi}\over h})^2)/ (n(m+1)-1)^2}
%$.
%\be v=
%{1\over 2}\sqrt{(3n-1)^2-(2n)^2({m_{\varphi}\over
%h})^2\over (mn+n-1)^2} \label{v2}\ee
%Note that there are three variables, and two constrained
%conditions which are from the requirements of the efolding number
%and scale invariant spectrum respectively, thus one variable can
%be left free.
which is used to replace $v$ in Eq.(\ref{kuk}), where the
subscript $m$ denotes the value of $v$ for the massive scalar
field. The scale invariance of spectrum requires $v_m=3/2$, thus
with Eq.(\ref{v2}), we can obtain \be
n(p+1)-1=\sqrt{\left(n-{1\over 3}\right)^2-\left({2n\over
3}\right)^2 \cdot\mu}. \label{v22}\ee Note that the term inside
sqrt in Eq.(\ref{v22}) should be larger than 0, which suggests
$\mu < (3/2 -1/(2n))^2$. For $0<n<1$, the range of $ (3/2
-1/(2n))^2 $ lies between $0$ and $\infty$. Thus to obtain enough
efolding number, $\mu <0 $ is generally required, which
corresponds to introduce a scalar field with negative mass term.
Substituting Eq.(\ref{v2}) into $\cal N$ obtained from
Eq.(\ref{caln2}), we can obtain \be {\cal
N}=\left(\sqrt{\left(n-{1\over 3}\right)^2-\left({2n\over
3}\right)^2 \cdot\mu}\right) \ln\left({h_i\over h_e}\right).
\label{nm} \ee
%From
%Eqs.(\ref{caln2}) and (\ref{v2}),
Thus one can see that to satisfy the requirements of observable
cosmology, for the decelerated expanding phase with $0<n<1$, the
enough efolding number may be obtained by properly selecting the
value of $\mu$ in the case with fixed $n$ in Eq.(\ref{nm}), while
when $n$ and $\mu$ are fixed by Eq.(\ref{nm}), the scale invariant
spectrum may be obtain by properly matching the value of $p$ in
Eq.(\ref{v2}). %For example, for the phase with $n=1/2$, we can
%take $\mu= -35/4 $, and thus have ${\cal N}=\ln(h_i/h_e)$, which
%can be leaded to enough efolding number very easily, and in the
%meantime from Eq.(\ref{v2}), we have $p =3$ which induces
%$v_m=3/2$, i.e. the scale invariant spectrum.
We plot Fig.3, in which for arbitrary $n$ in the region $0.4<n<1$,
in order to obtain enough efolding number, the $-\mu$ required
with respect to $n$ is given. We can see that for different value
of $n$, $-\mu$ lies in an acceptable region and in the meantime
$h_i/h_e$ is not required to be very large, which is from the case
with the massless scalar field. Note that both cases in Fig.3 can
not be implemented in the massless field, in which the value of
$h_i/h_e$ in Fig.3 is not so large to ensure enough efolding
number.

\begin{figure}[ht]
\begin{center}
\includegraphics[width=8cm]{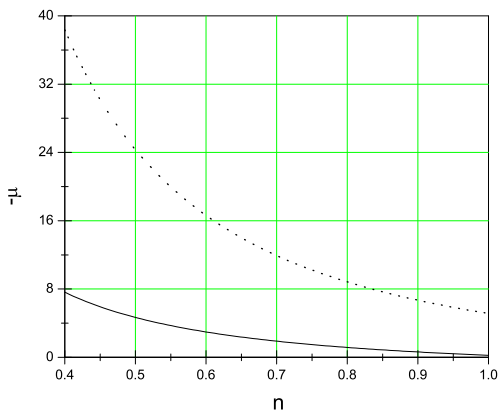}
\caption{The value of $-\mu$ with respect to $n$ in order to
obtain enough efolding number. The vertical axis is $-\mu$ and the
horizonal axis is $n$. The solid line corresponds to the case of
$h_i\sim 10^{-4}m_p$ and $h_e\sim$ Tev, and the dashed line is
that of $h_i\sim m_p$ and $h_e\sim 10^{-9}m_p$.}
\end{center}
\end{figure}

\section{summary and discussion}

The modification of the dispersion relation of the scalar field
brings a possibility generating the primordial perturbations in a
decelerated expanding background, however, we find that in order
to generate a nearly scale invariant spectrum, it will be slightly
difficult to obtain enough efolding number required by the
observable cosmology in a simple case. But when we consider more
general cases, e.g. the massive scalar field, the problem can be
relaxed. Thus though the conditions required look like slightly
special, it seems possible to seed the nearly scale invariant
primordial perturbations within a conventional evolution not
involving the inflation. These perturbations may be transferred to
the curvature perturbations at late time by some mechanisms, e.g.
as in Refs. \cite{DGZK, K}, thus may be interesting and
responsible for the structure formation of observable universe.
Note that current observations actually favor a red tilt spectrum
\cite{WMAP}, but not exact scale invariant one. However, this does
not pose any problem here, since we can always set any value of
$p$ in Eq.(\ref{v}) or (\ref{v2}) to obtain the $v$ or $v_m$
required by the red tilt of spectrum, even the blue tilt. In
principle there is not the generation of primordial gravitational
wave, since the seeding of scalar perturbation occurs during a
decelerated expansion, unless the sound speed of the gravitational
wave is also time dependent, as in the case of scalar
perturbation. Non-Gaussianity is expected to be small. However, we
still need to a detailed discussion on the gravitational wave and
non-Gaussianity in order to match the coming observable tests. In
addition, since the energy scale when the generating phase of
primordial perturbations ends may be very low to Tev, even BBN
scale, it is also interesting to study whether there are some
observable effect on e.g. baryogenesis. Thus it seems that many
significant issues related to this work remain. We expect to back
to these studies in the future.

Finally, it should be pointed out that such a decelerated
evolution of early universe can not solve all problems of standard
cosmology, as has been explained in the inflation models. For
example, here initially the homogeneity in the superHubble scale
%at the beginning of the generation of primordial perturbations
must be imposed. Be that as it may, however, this work
% our consideration, the homogeneity in superhorizon scale
%at the beginning of the generation of primordial perturbations
%must be imposed, which can be seen from Fig.2, and only be assumed
%here. But regardless, this work in some sense may be helpful to
displays an ``unnatural" but possible example seeding a
phenomenologically realistic spectrum of primordial perturbations
in a non-accelerated expanding background,
%enriches
%possibilities of seeding the primordial perturbations
which to some extent highlights the fact again that identifying
the origin of primordial perturbations may be a much more subtle
task than expected.
%deepening our exploring for the origin and nature of
%primordial perturbations from a different angle of view.
%In this sense the relevant issues in this note may be interesting
%to further study.

\textbf{Acknowledgments} The author would like to thank C.
Armendariz-Picon for discussions. This work is supported in part
by NNSFC under Grant Nos: 10405029, in part by the Scientific
Research Fund of GUCAS(NO.055101BM03), as well as in part by CAS
under Grant No: KJCX3-SYW-N2.

\end{document}